\documentclass[12pt ]{article}
\def\beq{\begin{equation}}
\def\eeq{\end{equation}}
\def\beqa{\begin{eqnarray}}
\def\eeqa{\end{eqnarray}}

\usepackage{epsfig}
\usepackage{rotate}

\begin{document}
\begin{titlepage}
\begin{center}
{\Huge Interacting strings on $AdS_3$}

\vspace*{1cm}

{\Large Gast\'on Giribet {\footnote {gaston@iafe.uba.ar}} 
and Carmen A. N\'u\~nez {\footnote {carmen@iafe.uba.ar}}} 

\vspace*{0,5cm}

Instituto de Astronom\'{\i}a y F\'{\i}sica del Espacio\\
IAFE - CONICET\\
C.C. 67 - Suc. 28, 1428 Buenos Aires, Argentina

\vspace*{1cm}


\begin{abstract}
We consider string theory on $AdS_3$ in terms of the Wakimoto free field
representation.
The scattering amplitudes for N unitary tachyons are analysed in the
factorization limit and the poles corresponding to the mass-shell
conditions for physical states are extracted. The vertex
operators for excited levels are obtained from the residues  
and their properties are examined. 
Negative norm states are found at the second mass level.
\end{abstract}

\end{center}
\begin{flushleft}
\end{flushleft}

\end{titlepage}

\section{Introduction}

String theory on three dimensional Anti-de Sitter spacetime has been
extensively studied in the past \cite{balog,pet,moh,hwang,bneme} as a toy
model to investigate the consistency of string propagation on background
fields. This is an interesting example since it provides a realization of
the Kac-Moody algebra of $SL(2,R)$, and is thus an exactly solvable model.
Renewed interest in the subject originated more recently after Maldacena's
conjecture \cite{malda} which suggests a duality between string theory
formulated on $AdS_D$ and a conformal field theory living on the $(D-1)$
dimensional boundary of $AdS$ spacetime. The case $D=3$ has attracted
special attention since the conjecture could be worked out in full detail
and explicitly verified \cite{gks,oo,ks} in this model.

However the structure of two dimensional non linear sigma models on non
compact spaces presents several problems. The quantum mechanical propagation
of strings on $AdS_3$ is not consistent. The Virasoro constraints are
not sufficient to decouple the negative norm states of the spectrum. Several
proposals have been advanced in the literature \cite
{pet,moh,hwang,bneme,bars,satoh} to render the theory unitary, but there is no
general agreement yet. A no ghost theorem has been proved for strings on $%
AdS_3$ at free level \cite{egp}. Unitarity is achieved by keeping only
states with quantum numbers $j$ bounded by the level of the algebra $k$ ($%
-1<j<k/2-1$). But this is not enough: a consistent string theory should
provide a mechanism to avoid the negative norm states at the interacting
level, similarly as in flat spacetime, i.e. non physical states must
decouple in physical processes. This implies that fusion rules should close
among ghost-free representations. Moreover, while the spectrum of the
compact current algebras can be truncated to the unitary representations
without spoiling modular invariance it is not clear that this can be done in
the non-compact case.

Furthermore it has been claimed that
 the $AdS/CFT$ correspondence suggests a relationship between the
conformal weights of the fields in the boundary conformal field theory, $h$
and the quantum number $j$ labeling the states of string theory in $AdS_3$ 
\cite{gks}. 
This assumption would lead to an inconsistency for 
the ground state tachyon belonging to the unitary
principal continuous representation $j\in -\frac 12+i\lambda $ \cite{oo}
since consistency of the spacetime $CFT$ requires $h$ to be real and 
thus it would darken the range of validity of the conjecture. 
It is possible that
the boundary $CFT$ gives a more comprehensive description of the fundamental
degrees of freedom of M-theory, but it is important to fill in the details
of the $AdS/CFT$ duality map in order to completly understand the
correspondence.

In standard perturbative string theory particles of various masses and spins
are exchanged in the different channels of a scattering amplitude. They
appear as singularities in the appropriate square momentum variable. One can
think of obtaining all the physical states of the spectrum of strings on $%
AdS_3$ and study their properties by analysing the scattering amplitudes of
the lowest energy particles in the theory. The purpose of this paper is to
address this question in the bosonic theory, starting from the correlation
functions of an arbitrary number of tachyons. Unlike string theory in
Minkowski spacetime, strings propagating on the non-compact $SL(2,R)$ group
manifold have an infinite degeneracy of states at a given mass level.
Nevertheless the scattering amplitudes exhibit several similarities with the
flat case. Singularities occur when the points where some of the external
vertex operators are attached coincide. The poles correspond to the mass
shell conditions for the intermediate states and the residues reproduce the
scattering amplitudes of the remaining tachyons with the exchanged states.
From this expression it is possible to extract the vertex operator
corresponding to the intermediate states, at arbitrary excited level.

In order to have a clear spacetime interpretation we consider the universal
covering of $SL(2,R)$ which is topologically $R^2\times R$, because time is
periodic on the group itself which is topologically $R^2 \times S^1$.
Moreover we work in the Wakimoto representation of the theory in terms of
free fields. In this formalism the string coupling constant depends on one
of the spacetime coordinates. Therefore our results are reliable near the
boundary of spacetime where the theory becomes weakly coupled. However since 
$AdS$ is a constant curvature spacetime it is reasonable to assume that the
spectrum of states of the theory does not depend on the region of spacetime
where it is located. In principle one should be able to find all the
possible states of the spectrum in this way (or at least all those that
couple to tachyons).

Conformal invariance imposes constraints on the vertex operators which are
automatically satisfied when this formalism is developed in flat spacetime
on arbitrary Riemann surfaces \cite{abin}. In the case under consideration
we have explicitly verified up to the second excited level that the vertex
operators obtained from factorization are automatically $CFT$ primaries.

This paper is organized as follows. In section 2 we review string theory on $%
AdS_3$. We summarize the $SL(2,R)$ WZW model in terms of the Wakimoto free
field representation and discuss the spectrum of string theory in this
background. In section 3 we present the tachyon vertex operator and the
formalism that has been developed in the literature to compute correlation
functions, following closely \cite{bb}. We perform the factorization of the $%
N$-tachyon scattering amplitude and obtain the poles corresponding to the
mass-shell condition for physical states. By analysing the residue we
reobtain the vertex operator for the intermediate tachyon state. In section
4 we study the higher order poles. We find the vertex operators for the
first and second excited levels and examine their properties.
At the first mass level all the states produced are unitary whereas 
negative norm states are found at the second mass level
even though the original amplitude involves only unitary external
tachyons. Finally the
conclusions in section 5 contain a discussion of the implications of
this result. As a by-product  
 the general form of the vertex operator for
an arbitrary excited level which can be expected from this particular
factorization process is given.

\section{Strings on $AdS_3$}

In this section we review some basic facts of string theory on $AdS_3$ in
order to introduce notation.

The $SL(2,R)$ Wess-Zumino-Witten model is given, in the Gauss parametrization,
by

\begin{equation}
S=k\int d^2z[\partial \phi \bar \partial \phi +\bar \partial \gamma \partial 
\bar \gamma e^{2\phi }]  \label{s1}
\end{equation}

This action describes strings propagating in three dimensional Anti-de
Sitter space with curvature $-\frac 2k$, metric 
\begin{equation}
ds^2=kd\phi ^2+ke^{2\phi }d\gamma d\bar \gamma  \label{gik}
\end{equation}
and background antisymmetric field 
\begin{equation}
B=ke^{2\phi }d\gamma \wedge d\bar \gamma  \label{bik}
\end{equation}

The boundary of $AdS_3$ is located at $\phi \rightarrow \infty $. Near this
region quantum effects can be treated perturbatively, the exponent in the
last term in (\ref{s1}) is renormalized and a linear dilaton in $\phi $ is
generated. Adding auxiliary fields $(\beta ,\bar \beta )$ and rescaling, the
action becomes \cite{dfk} 
\begin{equation}
S=\frac 1{4\pi }\int d^2z[\partial \phi \bar \partial \phi -\frac 2{\alpha
_{+}}R\phi +\beta \bar \partial \gamma +\bar \beta \partial \bar \gamma
-\beta \bar \beta e^{-\frac 2{\alpha _{+}}\phi }]  \label{s2}
\end{equation}
where $\alpha _{+}=\sqrt{2(k-2)}$. This action can be trusted for large
values of $\phi $.

This theory has a non-compact symmetry generated by currents $J^a(z)$ and $%
\bar J^a(\bar z)$, $a=1, 2, 3$. In the following we discuss the holomorphic
part of the theory because the same considerations apply to the
antiholomorphic part.

Expanding in a Laurent series, 
\begin{equation}
J^a(z) = \sum_{n=-\infty}^\infty J_n^a z^{-n-1}
\end{equation}
the coefficients $J_n^a$ satisfy a Kac-Moody algebra given by 
\begin{equation}
[J^a_n, J^b_m] = i f^{ab}_{~~~c} J^c_{m+n} - {\frac{k}{2}}g^{ab}n
\delta_{n+m,0}
\end{equation}
where the structure constants can be chosen as $f^{abc} = \epsilon^{abc}$
and the Cartan Killing metric is of the form $g^{ab} = diag(+1,+1,-1)$.

It is useful to introduce the complex basis $(J^\pm, J^3)$ where 
\begin{equation}
J^\pm = J^1 \pm i J^2 \quad , \quad (J^\pm)^{\dag} = J^\mp
\end{equation}
and the Kac Moody algebra takes the form 
\begin{equation}
[J^3_m, J^3_n] = -{\frac{k}{2}} m \delta_{m+n, 0}  \label{alg1}
\end{equation}
\begin{equation}
[J^3_m, J^\pm_n] = \pm J^\pm_{m+n}  \label{alg2}
\end{equation}
\begin{equation}
[J^+_m, J^-_n] = -2 J^3_{m+n} + km \delta_{m+n,0}  \label{algebra}
\end{equation}

This $SL(2,R)$ current algebra can be expressed in terms of the fields $%
(\phi, \beta,\gamma)$ using the Wakimoto representation \cite{waki} 
\begin{eqnarray}
J^{+}(z) &=&\beta (z)  \label{uaquimotito} \\
J^3(z) &=&-\beta (z)\gamma (z)-\frac{\alpha _{+}}2\partial \phi (z) 
\nonumber \\
J^{-}(z) &=&\beta (z)\gamma ^2(z)+\alpha _{+}\gamma (z)\partial \phi
(z)+k\partial \gamma (z)  \nonumber
\end{eqnarray}

Indeed, considering the free field propagators 
\begin{equation}
\left\langle \phi (z)\phi (w)\right\rangle =-\log (z-w)  \label{corelatuno}
\end{equation}
\begin{equation}
\left\langle \gamma (z)\beta (w)\right\rangle =-\frac 1{(z-w)}
\label{corelatdos}
\end{equation}
it is easy to verify that the OPEs of these currents realize a $SL(2,R)$
level $k$ Kac-Moody algebra, namely 
\begin{eqnarray}
J^{+}(z)J^{-}(w) &=&\frac k{(z-w)^2}-\frac 2{(z-w)}J^3(w)+...  \nonumber
\\
J^3(z)J^{\pm}(w) &=&\pm\frac 1{(z-w)}J^{\pm}(w)+...  \nonumber \\
J^3(z)J^3(w) &=&\frac{-k/2}{(z-w)^2}+...  \nonumber
\end{eqnarray}

The Sugawara construction leads to the following energy-momentum tensor 
\begin{equation}
T(z)=\beta \partial \gamma -\frac 12(\partial \phi )^2-\frac 1{\alpha _{+}}%
\partial ^2\phi  \label{jemmm}
\end{equation}
and hence the central charge of the theory is 
\begin{equation}
c=\frac{3k}{(k-2)}  \label{charche}
\end{equation}
This implies that $k=52/23$ when the theory is formulated on $AdS_3$. In
general there could be an internal compact space ${\cal N}$ ($i.e.$ the
theory lives on $AdS_3\times {\cal N}$), and in this case $k> 52/23$.

The spectrum of the theory can be obtained analysing the $SL(2,R)$ Kac-Moody
primary states. They are labeled by two quantum numbers $\left|
j,m\right\rangle $, where

\begin{eqnarray}
\Delta _o\left| j,m\right\rangle &=&j(j+1)\left| j,m\right\rangle
\label{guinpa} \\
J_o^3\left| j,m\right\rangle &=&m\left| j,m\right\rangle  \nonumber
\end{eqnarray}
and $\Delta _o=:-\frac 12(J_o^{+}J_o^{-}+J_o^{-}J_o^{+})+J_o^3J_o^3:$ is the
Casimir (the subindices refer to the zero-modes of the currents). The ground
states $\left| j,m\right\rangle $ satisfy 
\begin{eqnarray}
J_n^a\left| j,m\right\rangle &=&0 , \quad n>0  \label{wimpa}
\end{eqnarray}
The complete representation can be constructed by acting with the zero-modes
of the currents, namely

\begin{eqnarray}
J_o^{+}\left| j,m\right\rangle &=&(j-m)\left| j,m+1\right\rangle
\label{ginpa} \\
J_o^{-}\left| j,m\right\rangle &=&(-j-m)\left| j,m-1\right\rangle  \nonumber
\end{eqnarray}

It is possible to restrict the group representations by requiring
hermiticity and unitarity. Hermiticity implies that $J_o^3$ and $\Delta _o$
have a real spectrum, namely $\{m\in R$ $,$ $j\in R\}$ or $\{m\in R$ $,$ $%
j\in -\frac 12+iR\}$. Notice that in order to have a clear spacetime
interpretation we work on the universal covering of $SL(2,R)$. On the
group itself $m$ is quantized and restricted to be integer.
Unitarity at the base requires \cite{moh,bneme}

\[
m(m\pm 1)-j(j+1)\geq 0 
\]
The unitary representations of the universal covering of $SL(2,R)$ are
classified as follows: $a)$ the principal continuous series with $j = {\frac{%
1}{2}}\pm i\lambda$, $\lambda \in R$ and $m\in R$; $b)$ the 
discrete series, $D_j^\pm$ with highest and lowest weight  states having
$j=\pm m$, $m\in R$; $c)$ the
exceptional representation with $-{\frac{1}{2}}\le j < 0, m\in R$ and the
trivial representation with $j=m=0$.

The excited states are constructed by acting with $J_{-n}^a$, $n>0$ on the
ground state representation.

The string states must satisfy the Virasoro constraints $L_m\left| \Psi
\right\rangle =0$, $m>0$ and $L_o\left| \Psi \right\rangle =\left| \Psi
\right\rangle $. The last one implies 
\begin{equation}
-\frac{j(j+1)}{(k-2)}+L=1  \label{massshell}
\end{equation}
at excited level $L$. Notice that this expression is invariant under $%
j\rightarrow -j-1$ and therefore one can consider only states with $j>-{%
\frac{1}{2}}$ or $j<-{\frac{1}{2}}$.

Taking into account that the Casimir plays the role of mass squared
operator \cite{bneme}, the mass spectrum of the theory is 
\begin{equation}
M^2=\frac{(L-1)}2\alpha _{+}^2  \label{masita}
\end{equation}
Therefore, the ground state of the bosonic theory is a tachyon, the first
excited level contains massless states and there is an infinite tower of
massive states.

Notice that the on-shell tachyon of the theory living on $AdS_3$ has $j=-{%
\frac{1}{2}} \pm {\frac{i}{\sqrt {92}}} $ and thus belongs to the unitary
principal continuous series. If there is an internal compact space ${\cal N}$%
, eq. (\ref{massshell}) becomes 
\begin{equation}
-{\frac{j(j+1)}{{k-2}}} + L + \Delta = 1  \label{msip}
\end{equation}
where $\Delta$ is the contribution of the internal part.

\section{Interactions}

In this section we discuss string interactions on $AdS_3$. The aim is to
read the states of the spectrum and their vertex operators from the residues
of the poles in the combinations of quantum numbers corresponding to the
masses of the intermediate states.

As a starting point for the computation, the amplitude for the scattering of
physical particles has to be constructed. This procedure requires, as
initial data, the vertex operators of the external particles to be
scattered. The natural objects to start with are the lowest energy states,
i.e. tachyons. The tachyon vertex operator was originally constructed in
reference \cite{mor} by looking at the conformal dimensions and OPE with the
currents of fields in the $SL(2,R)_k$ $WZW$ theory, and it is given by 
\begin{equation}
V_T(j,m)=:\gamma ^{j-m}(z)\bar \gamma ^{j-\bar m}(\bar z)e^{\frac{2j}{\alpha
_{+}}\phi (z,\bar z)}:  \label{takion}
\end{equation}
where $m-\bar m\in Z$ is necessary for single valuedness. This vertex
operator should be supplemented with an appropriate internal part if the
theory is formulated on $AdS_3\times {\cal N}$.

Note that this is a different representation than the one introduced by
Teschner \cite{tecuno}. The operator (\ref{takion}) corresponds to the
leading large $\phi $ part of Teschner's representation, i.e. it is suitable
to work in the near boundary limit. 
The large $\phi $
approximation leads to incomplete results for correlation functions \cite
{oo,ks}. However it is sufficient for our purposes of extracting the short
distance singularities of the scattering amplitudes signaling the
on-mass-shell states of the theory. Indeed, it is reasonable to expect
that the spectrum of the theory will not 
depend on the region of spacetime where
the string is located (unless there were singularities).

We now review the formalism to construct the $N$-point function of tachyons.
This approach was developed in \cite{bb} and it is well suited to our
purposes. The starting point is the following scattering amplitude 
\begin{equation}
A_{m_1...m_N}^{j_1...j_N}=Vol[SL(2,C)]^{-1}\left\langle 
\mathop{\displaystyle \prod }
_{i=1}^N%
\displaystyle \int 
d^2z_iV_T(j_i,m_i)\right\rangle  \label{deene}
\end{equation}
where the average is performed with respect to the action (\ref{s2}). The
antiholomorphic dependence $j_i=\bar j_i$ and $m_i=\bar m_i$ is implicit in
this notation.

The integration over the zero-modes of the fields leads to a charge
conservation condition for non-vanishing amplitudes which can be achieved
through the inclusion of screening operators \cite{dotfat}. These operators
were constructed in reference \cite{mor} and can be represented as 
\begin{equation}
{\cal S}=\int d^2w\beta (w)\bar \beta (\bar w)e^{-\frac 2{\alpha _{+}}\phi
(w,\bar w)}  \label{screenico}
\end{equation}
Therefore the amplitude (\ref{deene}) takes the form 
\begin{equation}
A_{m_1...m_N}^{j_1...j_N}=Vol^{-1}[SL(2,C)]\left\langle 
\mathop{\displaystyle \prod }
_{i=1}^N%
\displaystyle \int 
d^2z_iV_T(j_i,m_i)\left( \int d^2w\beta \bar \beta e^{-\frac 2{\alpha _{+}}%
\phi }\right) ^s\right\rangle  \label{jkllkj}
\end{equation}
where the Gauss-Bonnet theorem implies, at the tree level 
\begin{equation}
s=\sum_{i=1}^Nj_i+1  \label{eidronth}
\end{equation}

Note that if the tachyons belong to the principal continuous series $s$ will
not be in general a positive integer. However, the prescription developed in 
\cite{dfk} for the equivalent situation in Liouville theory, consists in
assuming that the number of screenings is a positive integer and the
correlators are evaluated by analytic continuation.

The integration over the zero-modes of the $(\beta ,\gamma )$ system leads
to the following condition 
\begin{equation}
(number ~ of ~ \beta ~fields) - (number ~ of ~ \gamma ~ fields) =1
\label{bg}
\end{equation}
for non-vanishing correlation functions. In the case under consideration,
this equation is 
\begin{equation}
s=\sum_{i=1}^N(j_i-m_i)+1  \label{eidronthh}
\end{equation}

From (\ref{eidronth}) and (\ref{eidronthh}) the following conservation law
is obtained 
\begin{equation}
\sum_{i=1}^Nm_i=0  \label{keyfor}
\end{equation}

Therefore the $N$-point function for tachyons can be written as 
\begin{eqnarray*}
A_{m_1...m_N}^{j_1...j_N} &=&\frac 1{Vol[SL(2,C)]}%
\displaystyle \int 
\prod_{i=1}^Nd^2z_i%
\mathop{\displaystyle \prod }
_{n=1}^s%
\displaystyle \int 
d^2w_n\left\langle \prod_{i=1}^N\gamma _{(z_i)}^{j_i-m_i}\prod_{n=1}^s\beta
(w_n)\right\rangle \times  \label{inel} \\
&&\ \ \ \times \left\langle \prod_{i=1}^N\bar \gamma _{(\bar z_i)}^{j_i-
m_i}\prod_{n=1}^s\bar \beta (\bar w_n)\right\rangle \times \left\langle
\prod_{i=1}^Ne^{\frac 2{\alpha _{+}}j_i\phi (z_i,\bar z_i)}\prod_{n=1}^se^{-%
\frac 2{\alpha _{+}}\phi (w_n,\bar w_n)}\right\rangle
\end{eqnarray*}
Here $(z_i,\bar z_i)$ and $(w_n,\bar w_n)$ are the world-sheet coordinates
where the tachyonic and the screening vertex operators respectively are
inserted.

Using the propagators (\ref{corelatuno}) and (\ref{corelatdos}) this
amplitude becomes

\begin{eqnarray}
A_{m_1...m_N}^{j_1...j_N} &\sim &%
\displaystyle \int 
\mathop{\displaystyle \prod }
_{i=1}^Nd^2z_i%
\mathop{\displaystyle \prod }
_{r=1}^sd^2w_nC(z_i,w_n)\bar C(\bar z_i,\bar w_n)\prod_{i<k}^{}\left|
z_i-z_k\right| ^{-8j_ij_k/\alpha _{+}^2}\times  \nonumber \\
&&\ \ \times \prod_{i,n}\left| z_i-w_n\right| ^{8j_i/\alpha _{+}^2}\times
\prod_{n<m}^{}\left| w_n-w_m\right| ^{-8/\alpha _{+}^2}  \label{tyuyt}
\end{eqnarray}
where $C(z_i,w_n)$ and $\bar C(\bar z_i,\bar w_n)$ stand for the
contribution of the $(\beta ,\gamma )$ correlators (see eq.(\ref{losc}) 
below).

The corresponding expression in stardard string perturbation theory in
Minkowski spacetime exhibits some aspects of tree level unitarity. It has
poles for on-mass-shell states and the residue of each pole factorizes as
the product of tree amplitudes for subprocesses $1,2,3,...M,J\rightarrow
J,(M+1),...N$ as depicted in Fig.1. 

\bigskip
  \begin{figure}
  {\epsfig{file=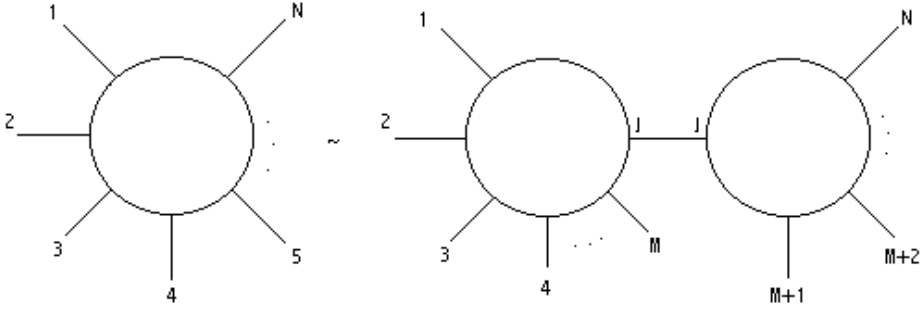, width=5.cm,angle=-90,
           bbllx=15, bblly=15, bburx=300, bbury=760,clip=} 
  }         
  \caption[]{Factorizacion of scattering amplitude}
  \end{figure}

The physical singularities of the amplitude (\ref{tyuyt}) can be extracted
by considering the region of integration where a subset of the $N$
tachyons collide to the same point on the world-sheet. Unitarity requires
also that the intermediate state $J$ has non-negative norm. We would like to
check whether these properties hold in string theory on $AdS_3$ when two
external vertex operators coincide. In this case, the factorization should
take the form 
\begin{equation}
A_{m_1...m_N}^{j_1...j_N}\sim \frac{\left\langle V_{T(j_1,m_1)}V_{T(j_2,m_2)}%
\tilde V_{T(j,m)}\right\rangle \times \left\langle
V_{T(j,m)}V_{T(j_3,m_3)}V_{T(j_4,m_4)}...V_{T(j_N,m_N)}\right\rangle }{-%
\frac{2j(j+1)}{\alpha _{+}^2}+L-1}  \label{en partes}
\end{equation}
where the conjugate vertex operator $\tilde V_{(j,m)}$ is $V_{(-j-1,-m)}$ 
\cite{mor}, $(j,m)$ are the quantum numbers of the intermediate state and $%
L\in Z$.

The conditions for non-vanishing residue in this expression are the
following 
\begin{equation}
s_1=j_1+j_2-j \qquad;\qquad m = m_1+m_2  \label{tra}
\end{equation}
\begin{equation}
s_2=%
\mathop{\displaystyle \sum }
_{i=3}^Nj_i+j+1\qquad;\qquad m+%
\mathop{\displaystyle \sum }
_{i=3}^Nm_i=0  \label{tre}
\end{equation}
where $s_1$ and $s_2$ are the number of screening operators in the $3$-point
and $(N-1)$-point functions respectively.

Let us consider the particular case $z_2\rightarrow z_1$ and take the $s_1$
contours to encircle $z_1$ and $z_2$ \cite{dotfat}. To isolate the
singularities corresponding to the intermediate channels perform the change
of variables: $z_1-z_2=\varepsilon ,$ $z_1-v_n=\varepsilon y_n,$ $%
v_n-z_2=\varepsilon (1-y_n)$, where we have renamed as $v_n$ the insertion
points of the $s_1$ contours.

In order to extract the explicit $\varepsilon $ dependence of the amplitude
it is convenient to write the functions $C(z_i,v_n,w_m)$ and $\bar C(\bar z%
_i,\bar v_n, \bar w_m)$ as 
\begin{eqnarray}
C(z_i,v_n, w_m) &=&\left\langle \gamma ^{j_1-m_1}(z_1)\gamma ^{j_2-m_2}(z_2)%
\mathop{\displaystyle \prod }
_{i=3}^N\gamma ^{j_i-m_i}(z_i)%
\mathop{\displaystyle \prod }
_{n=1}^{s_1}\beta (v_n)%
\mathop{\displaystyle \prod }
_{m=1}^{s_2}\beta (w_m)\right\rangle  \nonumber \\
\ &\sim&%
\mathop{\displaystyle \sum }
_{Perm(v_n)}%
\mathop{\displaystyle \sum }
_{k=0}^{s_1}\frac{(j_1-m_1)(j_1-m_1-1)...(j_1-m_1-k+1)}{%
(z_1-v_1)(z_1-v_2)...(z_1-v_k)}\times  \nonumber 
\end{eqnarray}
\begin{eqnarray}
&&~~~~~~~~~~~~~ \times \frac{(j_2-m_2)(j_2-m_2-1)..(j_2-m_2-s_1+k+1)}{%
(z_2-v_{k+1})...(z_2-v_{s_1})}\times  \nonumber \\
&&\ \times \left\langle \gamma ^{j_1-m_1-k}(z_1)\gamma ^{j_2-m_2-s_1+k}(z_2)%
\mathop{\displaystyle \prod }
_{i=3}^N\gamma ^{j_i-m_i}(z_i)%
\mathop{\displaystyle \prod }
_{m=1}^{s_2}\beta (w_m)\right\rangle +  \nonumber \\
&&\ \ +%
\mathop{\displaystyle \sum }
_{Perm(v_n)}%
\mathop{\displaystyle \sum }
_{k=0}^{s_1-1}\frac{(j_1-m_1)(j_1-m_1-1)...(j_1-m_1-k+1)}{%
(z_1-v_1)(z_1-v_2)...(z_1-v_k)}\times  \nonumber \\
&&\ \ \times \frac{(j_2-m_2)(j_2-m_2-1)..(j_2-m_2-s_1+k+2)}{%
(z_2-v_{k+1})...(z_2-v_{s_1-1})}%
\mathop{\displaystyle \sum }
_{i=3}^N\frac{j_i-m_i}{z_i-v_{s_1}}\times  \nonumber \\
&&\ \ \times \left\langle \gamma ^{j_1-m_1-k}(z_1)\gamma
^{j_2-m_2-s_1+k+1}(z_2)%
\mathop{\displaystyle \prod }
_{l\neq i}^{}\gamma ^{j_l-m_l}(z_l)\gamma ^{j_i-m_i-1}(z_i)%
\mathop{\displaystyle \prod }
_{m=1}^{s_2}\beta (w_m)\right\rangle +...\nonumber\\ 
\label{losc}
\end{eqnarray}
and similarly for $\bar C(\bar z_i,\bar v_n,\bar w_n)$. The sign $\sim$
stands for an irrelevant phase. The products $%
(j_1-m_1)(j_1-m_1-1)...(j_1-m_1-k+1)$ have to be understood as not
contributing for $k=0$ (similarly $(j_2-m_2)...(j_2-m_2-s_1+k+1)$ for $k=s_1$%
). The dots stand for lower order contractions between the fields inserted
at $z_1$ and $z_2$ and the $s_1$ screening operators. Note that these
functions can be written as a power series in $\varepsilon $ after
performing the change of variables and extracting the leading $\varepsilon
^{-s_1}$ divergence.

The amplitude becomes then formally 
\begin{eqnarray}
A_{m_1...m_N}^{j_1...j_N} &\sim &%
\displaystyle \int 
d^2\varepsilon \left| \varepsilon \right|
^{2s_1-[8j_1j_2-8s_1j_1-8j_2s_1+4s_1(s_1-1)]/\alpha _{+}^2-2s_1}\times 
\nonumber \\
&&\times 
\displaystyle \int 
d^2z_1\prod_{i=3}^N%
\displaystyle \int 
d^2z_i%
\mathop{\displaystyle \prod }
_{n=1}^{s_2}d^2w_n%
\mathop{\displaystyle \prod }
_{r=1}^{s_1}%
\displaystyle \int 
d^2y_r\left| \Phi (\varepsilon,z_1,z_i,y_r,w_n)\Psi (z_i,w_n)
\right|^2\nonumber\\
\label{yy}
\end{eqnarray}

The first term in the exponent of $\left| \varepsilon \right| $ comes from
the change of variables in the insertion points of the $s_1$ screening
operators whereas the last term cancelling it arises in the $\beta-\gamma$
system, eq. (\ref{losc}). The other terms in the exponent originate in the
contractions of the exponentials. The function $\Phi $ is a regular function
in the limit $\varepsilon \rightarrow 0$. It is convenient to write
separately the contribution to $\Phi$ from the exponentials ($E$) and from
the $\beta - \gamma$ system ($C$), i.e. $\Phi = E\times |C|^2$, where 
\begin{eqnarray}
E(\varepsilon, z_1, z_i, y_r, w_n) = \prod_{r=1}^{s_1}
&&|y_r|^{8j_1/\alpha_+^2} |1-y_r|^{8j_2/\alpha_+^2}
\prod_{r<t}|y_r-y_t|^{-8/\alpha_+^2}  \nonumber \\
\prod_{i=3}^N &&|z_1 - z_i|^{-8j_1j_i/\alpha_+^2} |z_1 - \varepsilon -
z_i|^{-8j_2 j_i/\alpha_+^2}  \nonumber \\
\prod_{m=1}^{s_2} &&|z_1 - w_m| ^{8j_1/\alpha_+^2} |z_1 - \varepsilon -
w_m|^{8j_2/\alpha_+^2}  \nonumber \\
\prod_{i=3}^N\prod_{r=1}^{s_1}&&|z_i - z_1 + \varepsilon y_r
|^{8j_i/\alpha_+^2} \prod_{r=1}^{s_1}\prod_{m=1}^{s_2} |z_1 - \varepsilon
y_r - w_m|^{-8/\alpha_+^2} \nonumber\\ 
\label{phi}
\end{eqnarray}
and 
\begin{eqnarray}
&&C(\varepsilon, z_1, z_i, y_r, w_n) \sim \mathop{\displaystyle \sum }
_{Perm(y_n)}%
\mathop{\displaystyle \sum }
_{k=0}^{s_1}\frac{(j_1-m_1)(j_1-m_1-1)...(j_1-m_1-k+1)}{y_1y_2...y_k}\times 
\nonumber \\
&&~~~~~~ \times \frac{(j_2-m_2)(j_2-m_2-1)...(j_2-m_2-s_1+k+1)}{%
(1-y_{k+1})(1-y_{k+2})...(1-y_{s_1})}\times  \nonumber \\
&&
\sum_{Perm(w_m)}{\large \{}{\large [ }{-\frac{(j_1-m_1-k)}{z_1-w_1}} - 
\frac{(j_2-m_2-s_1+k)} {(z_1-\varepsilon -w_1)} {\large ]} <\prod_{i=3}^N
\gamma^{j_i-m_i}(z_i) \prod_{m=2}^{s_2}\beta (w_m)> +  \nonumber \\
&&+ {\large [} \frac{(j_1-m_1-k)(j_1-m_1-k-1)}{(z_1 - w_1)(z_1-w_2)} + \frac{%
(j_2-m_2-s_1+k)(j_2-m_2-s_1+k-1)}{(z_1-\varepsilon-w_1)(z_1-\varepsilon -w_2)%
}  \nonumber \\
&&+ \frac{(j_1-m_1-k)(j_2-m_2-s_1+k)}{(z_1-w_1)(z_1-\varepsilon-w_2)} +\frac{%
(j_1-m_1-k)(j_2-m_2-s_1+k)}{(z_1-w_2)(z_1-\varepsilon-w_1)} {\large ]} 
\nonumber \\
&&~~~~~~~~~~~~~~~~~~~~~<\prod_{i=3}^N \gamma^{j_i-m_i}(z_i)
\prod_{m=3}^{s_2}\beta (w_m)> +... {\large \}} + ...
  \label{bg}
\end{eqnarray}
The dots inside the bracket in the last equation stand for terms involving
more contractions among the vertices at $z_1$ and $z_2$ and the $s_2$
screenings at $w_m$, whereas the dots at the end stand for lower order
contractions between the colliding vertices ($V_{T(j_1,m_1)}$ and
$V_{T(j_2,m_2)}$) and the $s_1$ screenings at $v_n$.

The function $\Psi $ in eq. (\ref{yy}) is independent of $\varepsilon $.

It is possible to Laurent expand $\Phi $ as 
\begin{equation}
\Phi =%
\mathop{\displaystyle \sum }
_{n,m,l,\bar l}\frac 1{n!m!}\varepsilon ^{n+l}\bar \varepsilon ^{m+\bar l%
}\partial ^n\bar \partial ^m\Phi _{l\bar l}{}_{\mid \varepsilon =\bar 
\varepsilon =0}  \nonumber
\end{equation}
where $\Phi _{l\bar l}$ denotes the contributions from terms in $%
C(z_i,v_n,w_m)$ and $\bar C(\bar z_i,\bar v_n, \bar w_m)$ where a number $l$
($\bar l$) of the $\beta$-fields $(\bar\beta)$ in the $s_1$ screenings are
not contracted with the $\gamma$-fields $(\bar\gamma)$ of the vertices at $%
z_1$ and $z_2$, but with the other vertices at $z_i$, $i=3,...N$.

Inserting this expansion in (\ref{yy}) and performing the integral over $%
\varepsilon $, the result is

\begin{eqnarray}
A_{m_1...m_N}^{j_1...j_N} &\sim &%
\mathop{\displaystyle \sum }
_{n,l}\frac{\Lambda ^{[-8j_1j_2+8s_1(j_1+j_2)-4s_1(s_1-1)]/\alpha
_{+}^2+2n+2l+2}}{-\frac 8{\alpha _{+}^2}j_1j_2+\frac 8{\alpha _{+}^2}%
s_1(j_1+j_2)-\frac 4{\alpha _{+}^2}s_1(s_1-1)+2n+2l+2}\times  \nonumber \\
&&\ \times 
\displaystyle \int 
d^2z_1%
\mathop{\displaystyle \prod }
_{t=1}^{s_2}%
\displaystyle \int 
d^2w_t\prod_{i=3}^N%
\displaystyle \int 
d^2z_i\prod_{r=1}^{s_1}%
\displaystyle \int 
d^2y_r\frac 1{(n!)^2}\partial ^n\Phi _l{}_{\mid \varepsilon =0}\Psi
(z_i,w_t)\times c.c. \nonumber \\
 \label{jkjkjk}
\end{eqnarray}
where $\Lambda $ is an infrared cut-off, irrelevant on the poles.

Let us analyse the pole structure of this expression, namely 
\begin{equation}
-\frac 4{\alpha _{+}^2}j_1j_2+\frac 4{\alpha _{+}^2}s_1(j_1+j_2)-\frac 2{%
\alpha_{+}^2}s_1(s_1-1)+1+n+l=0  \label{shouldg}
\end{equation}

Notice that for $n=l=0$ this is precisely the mass shell condition for a
tachyon with $j=j_1+j_2-s_1$, i.e. 
\begin{equation}
-\frac 2{\alpha _{+}^2}j(j+1)=-\frac 2{\alpha _{+}^2}%
(j_1+j_2-s_1)(j_1+j_2-s_1+1)=1  \label{shouldk}
\end{equation}
if $j_1$ and $j_2$ are on-mass-shell tachyons (i.e. $-\frac{2j_1(j_1+1)}{%
\alpha _{+}^2}=1$). As discussed in the previous section, this condition for
the ground state is satisfied only if $j= -\frac 12\pm \frac i{\sqrt {92}}.$

At this order, the residue reads

\begin{eqnarray}
A_{m_1...m_N}^{j_1...j_N}(\varepsilon =0) &=&%
\mathop{\displaystyle \prod }
_{r=1}^{s_1}%
\displaystyle \int 
d^2y_r%
\mathop{\displaystyle \prod }
_{r=1}^{s_1}\left| y_r\right| ^{8j_1/\alpha _{+}^2}\left| 1-y_r\right|
^{8j_2/\alpha _{+}^2}%
\mathop{\displaystyle \prod }
_{r<t}^{s_1}\left| y_r-y_t\right| ^{-8/\alpha _{+}^2}\times C^{\prime }(y_r)%
\bar C^{\prime }(\bar y_r)\times  \nonumber \\
&& \times 
\displaystyle \int 
d^2z_1%
\displaystyle \int 
\mathop{\displaystyle \prod }
_{i=3}^Nd^2z_i\prod_{n=1}^{s_2}%
\displaystyle \int 
d^2w_n\left| z_1-z_i\right| ^{-8(j_1+j_2-s_1)j_i/\alpha _{+}^2}\times 
\nonumber \\
&& \times \prod_{t=1}^{s_2}\left| z_1-w_t\right| ^{8(j_1+j_2-s_1)/\alpha
_{+}^2}%
\mathop{\displaystyle \prod }
_{3<i<k}^N\left| z_i-z_k\right| ^{-8j_ij_k/\alpha _{+}^2}%
\mathop{\displaystyle \prod }
_{i=3}^N%
\mathop{\displaystyle \prod }
_{t=1}^{s_2}\left| z_i-w_t\right| ^{8j_i/\alpha _{+}^2}\times  \nonumber \\
&& \times 
\mathop{\displaystyle \prod }
_{t<m}^{s_2}\left| w_t-w_m\right| ^{-8/\alpha _{+}^2}\times C^{\prime \prime
}(z_1,z_i,w_n)\bar C^{\prime \prime }(\bar z_1,\bar z_i,\bar w_n)
\label{popz}
\end{eqnarray}
where 
\begin{eqnarray*}
C^{\prime }(y_r) &=&%
\mathop{\displaystyle \sum }
_{Perm(y_n)}%
\mathop{\displaystyle \sum }
_{k=0}^{s_1}\frac{(j_1-m_1)(j_1-m_1-1)...(j_1-m_1-k+1)}{y_1y_2...y_k}\times
\\
&&\ \times \frac{(j_2-m_2)(j_2-m_2-1)...(j_2-m_2-s_1+k+1)}{%
(1-y_{k+1})(1-y_{k+2})...(1-y_{s_1})}
\end{eqnarray*}
and clearly from eq. (\ref{bg}) evaluated at $\varepsilon = 0$, 
\[
C^{\prime \prime }(z_1,z_i,w_n)=\left\langle \gamma
^{j_1-m_1+j_2-m_2-s_1}(z_1)%
\mathop{\displaystyle \prod }
_{i=3}^N\gamma ^{j_i-m_i}(z_i)%
\mathop{\displaystyle \prod }
_{m=1}^{s_2}\beta (w_m)\right\rangle 
\]

This can be easily interpreted as the product of a $3$-tachyon amplitude
(the first line in expression (\ref{popz})) 
\begin{equation}
\left\langle V_{T_{(j_1,m_1)}}(0)V_{T_{(j_2m_2)}}(1)\tilde V%
_{T_{(j,m)}}(\infty )%
\mathop{\displaystyle \prod }
_{r=1}^{s_1}{\cal S}(y_r)\right\rangle  \label{sallaw}
\end{equation}
times a $(N-1)$-tachyon amplitude 
\begin{eqnarray}
&&\left\langle \gamma _{(z_1)}^{(j_1+j_2-s_1)-m}\gamma
_{(z_3)}^{j_3-m_3}...\gamma _{(z_N)}^{j_N-m_N}\prod_{n=1}^{s_2}\beta
(w_n)\right\rangle \times  \label{wallas} \\
&&\times \left\langle \bar \gamma_{(\bar z_1)}^{(j_1+j_2-s_1)- m}\bar 
\gamma _{(\bar z_3)}^{j_3- m_3}...\bar \gamma _{(\bar z_N)}^{j_N- m%
_N}\prod_{r=1}^{s_2}\bar \beta (\bar w_n)\right\rangle \times  \nonumber \\
&&\times \left\langle e^{2(j_1+j_2-s_1)\phi (z_1,\bar z_1)/\alpha
_{+}}\prod_{i=3}^N e^{2j_i\phi (z_i,\bar z_i)/\alpha
_{+}}\prod_{n=1}^{s_2}e^{-2\phi (w_n,\bar w_n)/\alpha _{+}}\right\rangle 
\nonumber
\end{eqnarray}

Therefore, the tachyon vertex operator can be reconstructed, namely

\begin{equation}
V_{T(j,m)}(z,\bar z)
 ~ = ~ \gamma ^{j-m}(z)\bar \gamma ^{j-m}(\bar z)e^{\frac 2{\alpha
_{+}}j\phi (z,\bar z)}  \label{tachyk}
\end{equation}
with $j=j_1+j_2-s_1$ and $m=m_1+m_2$ (recall eq. (\ref{tra})). Since we
started with unitary external tachyons belonging to the principal continuous
representation, $m_1$ and $m_2$ can be arbitrary real numbers and then $m\in
R$. The intermediate tachyon satisfies the physical state conditions and its
norm is unitary by construction. It belongs to the principal continuous
series with quantum numbers $\left| j,m\right\rangle =\left|
j_1+j_2-s_1,m_1+m_2\right\rangle $.

Note that we could have started with external tachyons belonging to the
unitary discrete series. This is possible if a compact internal space ${\cal %
N}$ is allowed. In this case the conformal weight of the vertex operator
could be given fully by the internal piece, $i.e.$ $\Delta = 1$ in eq. (\ref
{msip}), and therefore the $AdS_3$ part could have $j_i(j_i+1) = 0$, namely $%
j_i=0$ or $-1$ (and for example $m_i=\pm j_i$ if the external tachyons
belong to the unitary discrete series) thus avoiding the continuous series.
Nevertheless string interactions still require an intermediate tachyon from
the principal continuous series, i.e. the factorization process discussed
above gives rise to an intermediate tachyon state belonging to the
continuous representation when the identity is taken in the internal part.
In fact, consider for example a free boson $X$ contributing a piece $%
\partial X$ to the external vertex operator (\ref{takion}) {\footnote{%
We thank J. Maldacena for suggesting this possibility.}}. The limit $%
z_2\rightarrow z_1$ produces a pole $\epsilon^{-2}$ from the internal part
which modifies eq. (\ref{shouldg}) to 
\begin{equation}
-\frac 4{\alpha _{+}^2}j_1j_2+\frac 4{\alpha _{+}^2}s_1(j_1+j_2)-\frac 2{%
\alpha_{+}^2}s_1(s_1-1)-1+n+l=0  \label{shouldg2}
\end{equation}
Thus the intermediate tachyon, $n=l=0$, has $j=j_1+j_2-s_1 \in -{\frac{1}{2}}%
\pm i \lambda$, $\lambda \in R$, $m\in Z$, and the identity in the internal
part when $j_1$ and $j_2$ belong to the discrete series ($i.e.$ $-\frac{2j_1%
}{\alpha_+^2}(j_1+1) = 0$ and, for example,  $m_1 = \pm j_1$).

\section{Excited states}

\subsection{The massless case}

Let us now consider the first order pole, corresponding to $n+l=1$. Both $%
n=1,l=0$ and $n=0,l=1$ contribute to (\ref{shouldg}) which can now be
written as 
\begin{equation}
-\frac 2{\alpha _{+}^2}j(j+1)=-\frac 2{\alpha _{+}^2}%
(j_1+j_2-s_1)(j_1+j_2-s_1+1)=0  \label{masyel}
\end{equation}
when $j_1$ and $j_2$ belong to the principal continuous representation.
Recalling the discussion at the end of last section, if $j_1$ and $j_2$
belong to the unitary discrete representations, either $D^-_j$ or $D^+_j$,
then the intermediate state should also satisfy eq. (\ref{masyel}) when we
consider the $\epsilon^{-2}$ contribution to the pole (corresponding to the
identity in the internal part of the intermediate state, $i.e.~~ \Delta = 0$).

Therefore there are two possible values for $j$, namely $j=0$ and $j=-1$. As
discussed initially in references \cite{pet,moh,hwang,bneme} and more
recently in \cite{egp}, some of all the possible states with these quantum
numbers have negative norm. In order to check the unitarity of the
interacting theory we will now examine whether ghost states are allowed in
the physical process under consideration.

Take the residue of the first order pole in eq. (\ref{jkjkjk}). Consider $%
n=1,l=0$ first. The derivative of the function $E$, eq. (\ref{phi}),
evaluated at $\varepsilon =0$ gives
\[
\partial _\varepsilon \Phi ={\large \{}%
\mathop{\displaystyle \sum }
_{i=3}^N\frac{4j_ij_2}{\alpha _{+}^2(z_1-z_i)}-%
\mathop{\displaystyle \sum }
_{n=1}^{s_2}\frac{4j_2}{\alpha _{+}^2(z_1-w_n)}-%
\mathop{\displaystyle \sum }
_{i=3}^N\sum_{r=1}^{s_1}\frac{4j_iy_r}{\alpha _{+}^2(z_1-z_i)}+
\]
\begin{equation}
+%
\mathop{\displaystyle \sum }
_{r=1}^{s_1}\sum_{n=1}^{s_2}\frac{4y_r}{\alpha _{+}^2(z_1-w_n)}{\large \}}%
\times c.c.%
\mathop{\displaystyle \sum }
\times \Phi (\varepsilon =0,z_1,z_i,y_r,w_n)\Psi (z_i,w_n)  \label{eluno}
\end{equation}

The first two sums in this expression can be clearly reproduced by the
scattering of $(N-2)$ tachyons with a term of the form $-\frac{2j_2}{\alpha
_{+}}\partial \phi (z_1)$. The last two double sums involve a different
integration over $y_r$. Notice that the integrals over $y_r$ in eq. (\ref
{popz}) can be evaluated using Dotsenko-Fateev (B.9) formula in the first
reference of \cite{dotfat}. In order to evaluate the integrals above which
include one extra $y_r$, note that the insertion $(1-2y_r)$ into the
Dotsenko-Fateev formula produces a vanishing result. Therefore, the $s_1$
terms containing $y_r$ in the numerator contribute to the residue one half
of the others and thus they can be reproduced by an operator $\frac{s_1}{%
\alpha _{+}}\partial \phi (z_1)$ scattering with the remaining $(N-2)$
tachyons. The vertex giving rise to (\ref{eluno}) is then 
\begin{equation}
-\frac 1{\alpha _{+}}(2j_2-s_1)\partial \phi (z)e^{2j\phi /\alpha _{+}}=-%
\frac j{\alpha _{+}}\partial \phi (z)e^{2j\phi /\alpha _{+}}  \label{partone}
\end{equation}
where we have taken into account that $j_2=j_1$ and $j=j_1+j_2-s_1$.

Another contribution to this residue arises when the derivative is taken on
the first term of $C(z_i,v_n,w_m)$ in eq. (\ref{losc}), $i.e.$ on the
coefficient of $\varepsilon^{-s_1}$. An inspection of eq. (\ref{bg}) shows
that it is not obvious how to factorize this expression after taking the $%
\varepsilon$-derivative. However the result greatly simplifies when
considering one of the external vertices which coincide at $z_1$, say $V_
{T(j_1, m_1)}$,
belonging to the highest weight representation of the unitary discrete
series $D_j^-$, $i.e.$ $j_1 = m_1$. In this case it is easy to reconstruct
the resulting expression by the scattering of $(N-2)$ tachyons at $z_i$ with
the operator 
\begin{equation}
-(j-m)\partial \gamma \gamma^{j-m-1}
\end{equation}

Finally, let us consider the contribution $n=0,l=1$ to the residue. The term 
$\Phi _1$ evaluated at $\varepsilon =0$ can be read from eq. (\ref{losc})

\[
\Phi _1\sim 
\mathop{\displaystyle \sum }
_{Perm(y_r)}%
\mathop{\displaystyle \sum }
_{k=0}^{s_1}\frac{(j_1-m_1)(j_1-m_1-1)...(j_1-m_1-k+1)
(j_2-m_2)...(j_2-m_2-s_1+k+2)}
{y_1y_2...y_k(1-y_{k+1})...(1-y_{s_1-1})}
\]
\[
\times
\mathop{\displaystyle \sum }
_{i=3}^N\frac{j_i-m_i}{(z_i-z_1)} \times
\]
\[
\times \mathop{\displaystyle \sum }
_{Perm(w_n)}%
\mathop{\displaystyle \sum }
_{t=0}^{s_2}\frac{(j_1-m_1-k)...(j_1-m_1-k-t+1)}{(z_1-w_1)...(z_1-w_t)} 
\times
\]
\begin{equation}
\times \frac{(j_2-m_2-s_1+k+1)(j_2-m_2-s_1-s_2+k+t+2)}{%
(z_1-w_{t+1})...(z_1-w_{s_2})}+...{\large \}}  \label{eppa}
\end{equation}
where the dots stand for contractions of some of the $s_2$ $\beta $-fields
in the screenings with the $\gamma $-fields in the $(N-2)$ remaining
tachyons at $z_i$. The last three lines of this expression can be reproduced
by an operator of the form $\beta (z_1)\gamma ^{j-m+1}(z_1)$
scattering with $(N-2)$ tachyons at $z_i$. In this case
the first line represents the three point function 
\begin{equation}
<\gamma ^{j_1-m_1}(0)\gamma ^{j_2-m_2}(1)\beta (\infty )\gamma
^{-j+m}(\infty )\prod_{r=1}^{s_1-1}\beta (y_r)>
\end{equation}
It is easy to verify that these correlators do not satisfy the charge
conservation conditions eqs. (\ref{tra}), (\ref{tre})
and thus they vanish. This result can be
generalized to arbitrary level. Notice that $\beta $ fields in the vertices
creating the intermediate states must be ``taken'' from the screening
operators, and thus these terms lead to violation of the conditions (\ref
{tra}), (\ref{tre}). Therefore we conclude that only the terms $\Phi _{l=0}$
contribute to the residues at all orders.

The vertex operator reproducing the residue of the first order pole is then
the following 
\begin{equation}
V^{(1)}_{(j,m)}(z, \bar z) = {\large [} - \frac {j}{\alpha_+} \partial \phi (z)
\gamma^{(j-m)} -
(j-m)\partial \gamma \gamma^{j-m-1}(z) {\large ]}
\times c.c.\times e^{2j\phi(z, \bar z)/\alpha_+}
\label{v1n}
\end{equation}
with $j=j_1+j_2-s_1$ and $m=m_1+m_2$. The coefficients of the operators
above play the role of polarization tensors. They are of course particular
ones obtained for the intermediate state at the first excited level produced
when two external tachyon insertions coincide. Moreover recall that this
particular operator is obtained when one of the external tachyons belongs to
the highest weight representation of the discrete series.

One could imagine more general polarizations, but the operatorial form of
the vertex operator in the Wakimoto representation is given by 
\begin{equation}
V=\left( A\partial \phi (z)\gamma ^{j-m}(z)+B\gamma ^{j-m-1}(z)\partial
\gamma (z)\right) \times c.c.\times e^{\frac 2{\alpha _{+}}j\phi (z,\bar z)}
\label{v1}
\end{equation}

It can be verified that this operator has conformal weight 1 when $-\frac{2j%
}{\alpha_+}(j+1)=0$, which is the mass-shell condition, for any $A$ and $B$.
The OPE with the energy-momentum tensor contains a cubic pole whose residue
(corresponding to $L_1|V>$)
is 
\[
-\frac{2A}{\alpha _{+}}(j+1) 
\]
and thus it automatically vanishes for the vertex (\ref{v1n})
obtained from the factorization.

In order to compute the norm of the state created by this vertex notice that
a level one state can be written in the operatorial formalism as 
\begin{equation}
V^{(1)} = a_0 J_{-1}^3|j,m> + a_+ J_{-1}^+ |j,m-1> + a_- J_{-1}^-|j,m+1>
\end{equation}
The norm of this general state can be computed using the commutators 
(\ref{alg1})-(\ref
{algebra}) and it is 
\begin{equation}
N = - \frac k2 a_0^2 + a_+^2 [k+2(m-1)] + a_-^2 [k-2(m+1)] + a_0 a_- (1+2m)
+ a_0 a_+ (1-2m)
\end{equation}

In the case of a vertex of the form (\ref{v1}) the coefficients are 
\begin{eqnarray}
a_0 = 2\frac{B \alpha_++A(2m+2-k)}{\alpha_+(k-2-2j)}  \nonumber \\
a_+ = -\frac A{\alpha_+}, \qquad a_- = \frac A{\alpha_+} + \frac{a_0}2.
\end{eqnarray}
and the norm is $N = A^2$. For the particular case of the state obtained in
the factorization, eq. (\ref{v1n}), it is then
\begin{equation}
N = \frac{j^2}{\alpha_+^2}
\end{equation}
Therefore it is a non-negative norm state since $j = 0$ or $-1$. Notice that
the norm is independent of $m$, thus this result holds for arbitrary
external tachyons belonging to $D^\pm_j$ as long as one of the coinciding
vertices creates a heighest weight state. Moreover the norm is
non-negative independently of $k$ (recall that $k\ge 52/23$) .

\subsection{Second excited level}

The pole corresponding to the second excited level, $n=2$, is given by 
\begin{equation}
-\frac 2{\alpha_+^2} j (j+1) + 2 = 1 \label{n=2msc}
\end{equation}
again with $j=j_1+j_2-s_1$ and $m=m_1+m_2$, therefore in this case $j=-\frac 
12\pm\sqrt{k-\frac 74}\in R$.  In order to extract the
vertex operator for this state we have to perform two derivatives on the
function $\Phi$.

The $\beta$-$\gamma$ system can be easily analysed if we consider, as in the
previous subsection, one of the external tachyons belonging to the highest
weight representation. One of the contributions to the residue arises when
one of the derivatives is taken over $E(\varepsilon, z_1, z_i, w_n)$ 
and the other one over $C(\varepsilon, z_1, z_i, w_n)$. This
can be reproduced with the following operator 
\begin{equation}
\frac {2j}{\alpha_+}(j-m)\partial \phi \partial \gamma \gamma^{(j-m-1)}
e^{2j\phi/\alpha_+}
\end{equation}
which is basically twice the product of the two terms in $V^{(1)}$.

Repeating the steps performed in the massless case, when the two derivatives
act on $C(z_i, w_n)$ (eq. (\ref{bg})) we obtain 
\begin{equation}
(j-m)\partial^2 \gamma \gamma^{j-m-1} + (j-m) (j-m-1)(\partial \gamma)^2
\gamma^{j-m-2}
\end{equation}

Finally, when the two derivatives act on $E(z_1,z_i,w_n)$ we encounter a
technical difficulty arising from the fact that there appear $\sum_r y_r^2$
factors in the integrands of Dotsenko-Fateev 
formula, which we are unable to handle.
Nevertheless, the operatorial form of the terms reproducing this part of the
residue can be read clearly, and it is 
\begin{equation}
(A \partial^2 \phi(z) + B \partial \phi(z) \partial \phi(z) )\gamma^{(j-m)}
\times ~c.c.~\times ~ e^{2j\phi(z,\bar z)/\alpha_+}
\end{equation}

The complete vertex operator at the second excited level is then, 
\begin{eqnarray}
V^{(2)}_{(j,m)}(z,\bar z)
 &=& {\large \{}(A \partial^2 \phi(z) + B \partial \phi(z)
\partial \phi(z) ) \gamma^{(j-m)} + (j-m)\partial^2 \gamma \gamma^{j-m-1} + 
\nonumber \\
&&+(j-m) (j-m-1)(\partial \gamma)^2 \gamma^{j-m-2} + 2\frac j{\alpha_+}%
(j-m)\partial \phi \partial \gamma \gamma^{(j-m-1)} {\large \}}\nonumber\\
&&\times ~ c.c.~\times ~e^{2j\phi(z,\bar z)/\alpha_+}
\label{secmass}
\end{eqnarray}

It is easy to verify that the OPE of this expression with the energy momentum
tensor corresponds to a weight one primary field if the undetermined
coefficients $A$ and $B$ satisfy the constraints 
\begin{eqnarray}
\frac 6 {\alpha_+} A + \frac {4 j}{\alpha_+} A + B = 0  \nonumber \\
A - \frac 2{\alpha_+} B (j+1) = 0  \label{psc}
\end{eqnarray}
corresponding to $L_2|phys> = 0$ and $L_1|phys>=0$ respectively.
These equations are compatible with the mass-shell condition
(\ref{n=2msc}) only if
$A=B=0$. 
The norm of the state
created by this vertex operator
can be computed following the same steps as in the
first excited level, namely one can write
it as a linear combination of the
currents $J_{-2}^a$ and  $J_{-1}^aJ_{-1}^b$
and use the commutators (\ref{algebra}).
 The result is a long expression
depending on $k$ and $m$. 
We have 
plotted 
in Figure 2 the norm obtained for the states $j=-{1\over 2}-\sqrt{k-7/4}$,
$m$ negative. Similar figures are obtained for $m$ positive and for
$j=-{1\over 2}+\sqrt{k-7/4}$, $m$ both positive and negative.

  \begin{figure}
  {\epsfig{file=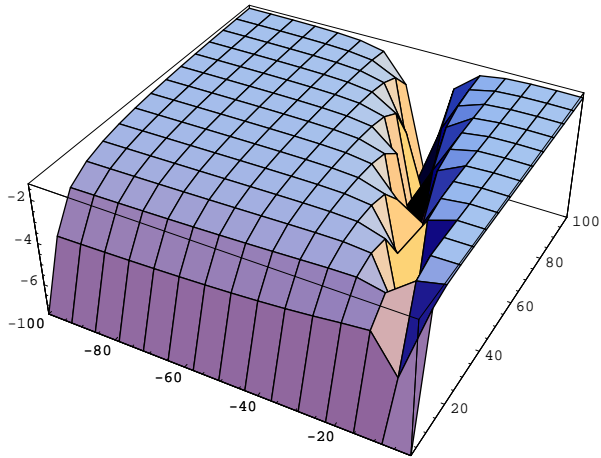, width=10.cm,
           bbllx=0, bblly=0, bburx=480, bbury=343,clip=} 
  }         
  \caption[]{Norm of the states $j=-{1\over 2}-\sqrt{k-7/4}$, 
as a function of
 $m< 0$ and $k$.}
  \end{figure}

The states have either negative or zero norm. We must
stress that this 
feature arises for the particular states (\ref{secmass}) produced from 
the scattering of one highest
weight  and one arbitrary 
tachyon both belonging to the discrete representation.

\section{Conclusions}

We have studied the factorization properties of scattering amplitudes of $N$%
-tachyons in the Wakimoto representation of bosonic string theory on $AdS_3$%
. The pole structure appearing when two vertex operators coincide on the
world-sheet, reproduces the mass-shell conditions for arbitrarily excited
level.

Using the expression for the on-shell three point function computed
in reference \cite{bb} it is easy to see that the residues are
different from zero.
By analysing them we were able to obtain the vertex operators
creating physical states and study their properties.

The tachyon vertex was reproduced for the ground state. In this case, the
mass-shell condition implies that the state produced by the factorization
belongs to the principal continuous representation. This is so even though
it is possible to avoid this series from the beginning by considering a non
trivial contribution to the external states from the internal space.

The vertex operator creating a massless state was found in the particular
case when one of the external tachyons which collide to the same point on
the world-sheet belongs to the heighest weight representation. The norm of
the first excited state produced in this way was computed. It is
non-negative independently of the quantum number $m$ and of the level of the
algebra $k$ (recall $k\ge 52/23$ in these models). It is interesting
to note that the
operatorial form (\ref{v1}) is, to this order, the most general 
one that can couple
to tachyons
(the coefficients could be more general). 
In fact, terms containing $\beta$ fields cannot be produced in
this way since they lead to violation of the charge conservation conditions
which are necessary to obtain a non vanishing result.

Finally we have found the general form of the vertex operator producing a
state of the second excited level. Even though we cannot obtain the
full vertex operator directly from the factorization we can determine
it from the conditions it has to satisfy to be a weight one primary
field.
The norms of these particular states
turn out to be non-positive.
Therefore, although we started with unitary external states the
 interactions introduce ghosts into the theory.
This is in accordance with the results found in reference \cite{petro} where
modular transformations of characters for generic values of $k$
lead to violation of the unitarity bound of string theory on
$AdS_3$ with a finite number of mass levels.

The procedure which we have developed allows to obtain the general form of
the vertex operators for any mass level. At excited level $L$ it can be
written as the following sum

\begin{equation}
V^{(L)}_{(j,m)}(z,\bar z) = \sum_{k,l,n,p = 0}^L A_{klnp}
\phi^{-\delta_{n,0}} \partial^n \phi (z)
(\partial \phi (z))^k \partial^l \gamma(z) (\partial \gamma (z))^p
\gamma^{j-m-p-1} \times c.c.~ e^{2j\phi/\alpha_+}
\end{equation}
where $k+l+n+p = L$ and the factor $\phi^{-\delta_{n,0}}$ is included to 
cancel the contribution $\partial^0 \phi$.
\bigskip

To conclude, in this paper we have proposed 
  a method and developed the corresponding  formalism to study
the unitarity of interacting string theory on $AdS_3$.
Even restricting the external states to those satisfying the unitarity bound
the interactions produce negative norm states. These results suggest that 
the proposed cut off over the values of $j$
should be reconsidered having into account the role of interactions. 

Many interesting questions remain. In particular,
it is necessary to find a 
physical mechanism to decouple the ghosts (such as the GSO projection in
the superstring theory decouples the tachyons). Otherwise 
the negative norm states
 should be interpreted in physical terms as arising from
some kind of instability.
Moreover it would be interesting to extend this analysis to the
supersymmetric case.

\bigskip
\noindent
{\bf {Acknowledgements}}

We would like to thank J. Maldacena, H. Ooguri and J. Russo for many useful
discussions and suggestions. This work was supported by CONICET, Argentina,
PIP 0873/98.


\end{document}